\documentclass[letterpaper,english,pre,intlimits,preprint]{revtex4}
\usepackage{times}
\usepackage[T1]{fontenc}
\usepackage[latin1]{inputenc}
\usepackage{amsmath}
\usepackage{graphicx}
\usepackage{amssymb}

\makeatletter

\providecommand{\tabularnewline}{\\}

\usepackage{babel}
\makeatother
\begin{document}

\title{Forecast and Control of Epidemics in a Globalized World}

\author{L. Hufnagel}

\author{D. Brockmann}

\author{T. Geisel}

\affiliation{Max-Planck-Institut für Strömungsforschung, Bunsenstrasse 10, 37073
Göttingen, Germany\\
and Kavli Institute for Theoretical Physics, University of California
Santa Barbara, CA 93106, USA}

\begin{abstract}
The rapid worldwide spread of the severe acute respiratory syndrome
(SARS) demonstrated the potential threat an infectious disease poses
in a closely interconnected and interdependent world. Here we introduce
a probabilistic model which describes the worldwide spreading of infectious
diseases and demonstrate that a forecast of the geographical spread
of epidemics is indeed possible. It combines a stochastic local infection
dynamics between individuals with stochastic transport in a worldwide
network which takes into account the national and international civil
aviation traffic. Our simulations of the SARS outbreak are in suprisingly
good agreement with published case reports. We show that the high
degree of predictability is caused by the strong heterogeneity of
the network. Our model can be used to predict the worldwide spreading
of future infectious diseases and to identify endangered regions in
advance. The performance of different control strategies is analyzed
and our simulations show that a quick and focused reaction is essential
to inhibit the global spreading of epidemics. 
\end{abstract}
\maketitle

\newcommand{\diff}[1]{\text{d}#1}

\newcommand{\pdt}{\partial_{t}\,}

\newcommand{\ablt}[1]{\diff{#1}/\diff{t}}

\newcommand{\zvec}[1]{\mathbf{#1}}

\section{Introduction}

\begin{flushleft}%
\begin{figure}
\includegraphics[%
  width=1.0\columnwidth]{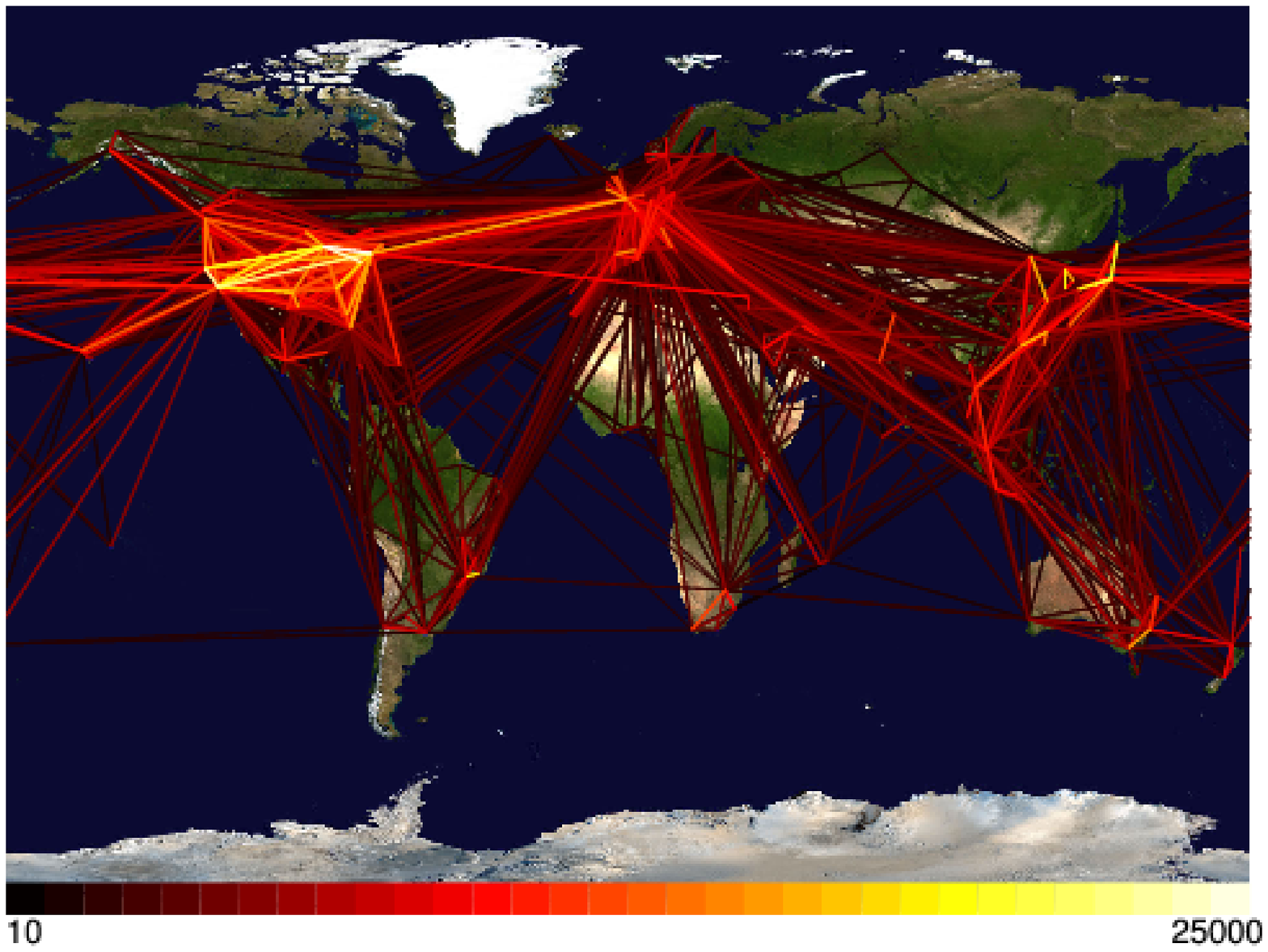}

\caption{\label{cap:fig1}\textsf{Global aviation network. A geographical
representation of the civil aviation traffic among the $500$ largest
international airports in over $100$ different countries is shown.
Each line represents a direct connection between airports. The color
encodes the number of passengers per day (see color code at the bottom)
traveling between two airports. The network accounts for more than
$95\%$ of the international civil aviation traffic. For each pair
$(i,j)$ of airports, we checked all flights departing from airport
$j$ and arriving at airport $i$. The amount of passengers carried
by a specific flight within one week can be estimated by the size
of the aircraft (We used manufacturer capacity information on over
$150$ different aircraft types) times the number of days the flight
operates in one week. The sum of all flights yields the passengers
per week, i.e. $M_{ij}$ in Eq.~(\ref{eq:rate}). We computed the
total passenger capacity $\sum M_{ij}$ of each airport $j$ per week
and found very good agreement with independently obtained airport
capacities.}}
\end{figure}
\end{flushleft}

The application of mathematical modeling to the spread of epidemics
has a long history and was initiated by Daniel Bernoulli's work on
the effect of cow-pox inoculation on the spread of smallpox in 1760\cite{bernoulli}.
Most studies concentrate on the local temporal development of diseases
and epidemics. Their geographical spread is less well understood,
although important progress has been achieved in a number of case
studies~\cite{keeling_813:2001,smith_3668:2002,keeling_136:2003}.
The key question and difficulty is how to include spatial effects
and to quantify the dispersal of individuals. This problem has been
studied with some effort in various ecological systems, for instance
in plant dispersal by seeds~\cite{bull:2002}. Today's volume, speed
and non-locality of human travel~(Fig.~\ref{cap:fig1}) and the
rapid worldwide spread of SARS~(Fig.~\ref{cap:fig2}) demonstrate
that modern epidemics cannot be accounted for by local diffusion models
which are only applicable as long as the mean distance traveled by
individuals is small compared to geographical extents. These local
reaction-diffusion models generically lead to epidemic wavefronts,
which were observed for example in the geotemporal spread of the Black
Death in Europe from 1347-50~\cite{Murray93,langer,noble74,grenfell01,mollison}. 

Here we focus on mechanisms of the worldwide spreading of infectious
diseases. Our model consists of two parts: a local infection dynamics
and the global traveling dynamics of individuals similar to the models
investigated in~\cite{ravchev_3:1985}. However, both constituents
of our model are treated on a stochastic level, taking full account
of fluctuations of disease transmission, latency and recovery on one
hand, and fluctuations of the geographical dispersal of individuals
on the other. Furthermore we incorporate nearly the entire civil aviation
network.

\section{Local Infection Dynamics}

In the standard deterministic SIR model for infectious diseases, a
population with $N$ individuals is categorized according to its infection
status: susceptibles ($S$), infectious ($I$) or recovered and immune
($R$)\cite{Murray93,Anderson91}. The dynamics which specifies the
flow among these categories is given by\begin{equation}
\ablt{s}=-\alpha\, s\, j,\qquad\ablt{j}=\alpha\, s\, j-\beta\, j\quad,\label{eq:localdynamics}\end{equation}
where $s=S/N$ and $j=I/N$ denote the relative number of susceptibles
and infecteds, respectively. The relative number of recovered individuals
$r=R/N$ is obtained by conservation of the entire population, i.e.
$r(t)=1-j(t)-s(t)$, and $\tau=\beta^{-1}$ is the average infectious
period. The key quantity describing the infection is the basic reproduction
number $\rho_{0}=\alpha/\beta$, which is the average number of secondary
infections transmitted by an infectious individual in an otherwise
uninfected population. If $\rho_{0}>1$ and the initial relative number
of susceptibles is greater than a critical value $s_{c}=1/\rho_{0}$
an epidemic develops ($\ablt{j}>0$). As the number of infected individuals
increases, the fraction of susceptibles $s$ decreases and thus the
number of contacts of infected individuals withfig susceptibles decreases
until $s=s_{c}$ when the epidemic reaches its maximum and subsequently
decays. 

The above SIR model incorporates the underlying mechanism of transmission
and recovery dynamics and has been able to account for experimental
data in a number of cases. However, transmission of and recovery from
an infection are intrinsically stochastic processes and the deterministic
SIR model does not account for fluctuations. These fluctuations are
particularly important at the beginning of an epidemic when the number
of infecteds is very small. 

In this regime a probabilistic description must be used. Schematically
the stochastic infection dynamics is given by

\begin{equation}
S+I\xrightarrow{\alpha}2\, I,\qquad I\xrightarrow{\beta}\emptyset\quad.\label{eq:sir_reaction}\end{equation}
 The first reaction reflects the fact that an encounter of an infected
individual with a susceptible results in two infecteds at a probability
rate $\alpha$, the second indicates that infecteds are removed (recover)
at a rate $\beta$ and effectively disappear from the population.
The quantity of interest is the probability $p(S,I;t)$ of finding
a number $S$ of susceptibles and $I$ infecteds in a population of
size $N$ at time $t$. Assuming that the process is Markovian on
the relevant time scales, the dynamics of this probability is governed
by the master equation~\cite{gardi:1985}\begin{eqnarray}
\pdt p(S,I;t) & = & \frac{\alpha}{N}(S+1)\,(I-1)\, p(S+1,I-1;t)+\beta\,(I+1)\, p(S,I+1;t)\nonumber \\
 & - & \left(\frac{\alpha}{N}SI\, p-\beta\, I\right)p(S,I;t)\quad.\label{eq:master2}\end{eqnarray}
In addition to this dynamics one must specify the initial condition
$p(S,I;\, t=t_{0})$ which is typically assumed to be a small but
fixed number of infecteds $I_{0}$, i.e. $p(S,I;t=t_{0})=\delta_{I,I_{0}}\delta_{S,N-I_{0}}$.

The relation of the probabilistic master equation~(\ref{eq:master2})
to the deterministic SIR-model~(\ref{eq:localdynamics}) can be made
in the limit of a large but finite population, i.e. $N\gg1$. In this
limit one can approximate the master equation by a Fokker-Planck equation
by means of an expansion in terms of conditional moments (Kramers-Moyal
expansion~\cite{gardi:1985}), see the supplement material. The associated
description in terms of stochastic Langevin equations reads\begin{eqnarray}
\ablt{s} & = & -\alpha\, s\, j+\frac{1}{\sqrt{N}}\sqrt{\alpha\, s\, j}\xi_{1}(t)\label{eq:final_lvs}\\
\ablt{j} & = & \alpha s\, j-\beta\, j-\frac{1}{\sqrt{N}}\sqrt{\alpha\, s\, j}\xi_{1}(t)+\frac{1}{\sqrt{N}}\sqrt{\beta\, j}\xi_{2}(t)\label{eq:final_lvj}\end{eqnarray}
Here, the independent Gaussian white noise forces $\xi_{1}(t)$ and
$\xi_{2}(t)$ reflect the fluctuations of transmission and recovery,
respectively. Note that the magnitude of the fluctuations are $\propto1/\sqrt{N}$
and disappear in the limit $N\rightarrow\infty$ in which case Eqs.~(\ref{eq:localdynamics})
are recovered. However, for large but finite $N$ a crucial difference
is apparent: Eqs.~(\ref{eq:final_lvs}) and~(\ref{eq:final_lvj})
contain fluctuating forces and $N$ is a parameter of the system.
A careful analysis shows that even for very large populations (i.e.
$N\gg1$) fluctuations play a prominent role in the initial phase
of an epidemic outbreak and cannot be neglected. For instance even
when $\rho_{0}>0$, a small initial number of infecteds in a population
may no necessarily lead to an outbreak which cannot be accounted for
by the deterministic model.

\section{Dispersal on the Aviation Network}

As individuals travel around the world, the disease may spread from
one place to another. In order to quantify the traveling behavior
of individuals, we have analyzed all national and international civil
flights among the $500$ largest airports by passenger capacity. This
analysis yields the global aviation network shown in Fig.~\ref{cap:fig1},
further details of the data collection is compiled in the supplement
material. The strength of a connection between two airports is given
by the passengers capacity, i.e. the number of passengers that travel
this route per day. 

We incorporate the global dispersion of individuals into our model
by dividing the population into $M$ local urban populations labeled
$i$ containing $N_{i}$ individuals. For each $i$ the number of
susceptibles, infecteds individuals is given by $S_{i}$ and $I_{i}$,
respectively. In each urban area the infection dynamics is governed
by the master equation~(\ref{eq:master2}). 

Stochastic dispersal of individuals is defined by a matrix $\gamma_{ij}$
of transition probability rates between populations\begin{equation}
S_{i}\xrightarrow{\gamma_{ij}}S_{j}\qquad I_{i}\xrightarrow{\gamma_{ij}}I_{j},\qquad i,j=1,...,M\quad,\label{eq:pups}\end{equation}
where $\gamma_{ii}=0$. Along the same lines as presented above one
can formulate a master equation for the pair of vectors $\zvec{X}=\left\{ S_{1},I_{1},...,S_{M},I_{M}\right\} $
which defines the stochastic state of the system. This master equation
is provided explicitely in supplement material. 

In order to account for the global spread of an epidemic via the aviation
network one needs to specify the matrix $\gamma_{ij}$. Since the
global exchange of individuals between urban areas is carried out
by airborne travel one can estimate the probability rate matrix $\gamma_{ij}$
by t. We assume that an individual remains in urban area for some
time before traveling to another region. A flight $j\rightarrow i$
is chosen according to the weights

\begin{equation}
w_{ij}=M_{ij}/\sum_{i}M_{ij}\quad.\label{eq:rate}\end{equation}
where $M_{ij}$ is the number of passengers per unit time that depart
from an airport in region $j$ and arrive at airport in region $i$.
The matrix $w$ accounts for the overall connectivity of the aviation
network as well as for the heterogeneity in the strength of the connections.
Denoting the typical time period individuals remain at $i$ by $\tau_{i}$
the matrix $\gamma_{ij}$ is expressed in terms of $w_{ij}$ according
to $\gamma_{ij}=w_{ij}/\tau_{j}$. If we assume that each airport
is surrounded by a catchment area with a population $N_{i}$ the typical
time individuals remain at $i$ is given by $\tau_{i}=N_{i}/\sum_{j}M_{ji}$.
If the capacity of airport $i$ reflects the need of the associated
catchment area (i.e. $N_{i}\propto\sum_{j}M_{ji}$), the waiting times
$\tau_{i}$ are identical for all $i$, i.e. $\tau_{i}=\tau=\gamma^{-1}$
which implies $\gamma_{ij}=\gamma\, w_{ij}$. In our model the global
rate $\gamma$ is a free parameter. In order to verify the its validity,
we apply our model to the SARS outbreak. The rate $\gamma$ can be
computed from the ratio of the number of infected individuals in Hong
Kong to the number of infected individuals outside Hong Kong, which
is provided by the WHO data. For the local infection dynamics we use
a simple extension of the above stochastic SIR model: The categories
$S$, $I$ and $R$ are completed by a category $L$ of latent individuals
which have been infected but are not infectious yet themselves, accounting
for the latency of the disease. In our simulations individuals remain
in the latent or infectious stage for periods drawn from the delay
distribution provided in Fig.~2 in~\cite{donnelly:1762_2003}. In
our simulation we chose random infection times the distribution of
which is known for SARS~\cite{donnelly:1762_2003}. In a realistic
simulation the basic reproduction number $\rho_{0}$ cannot be assumed
to be constant over time. Successful control measures, for instance,
generally decrease $\rho_{0}$. We chose a time dependent $\rho_{0}(t)$
as provided by Refs.~\cite{riley03,lipsitch03}.

\section{Results of Simulations}

\begin{figure}
\begin{center}\includegraphics[%
  width=0.70\columnwidth]{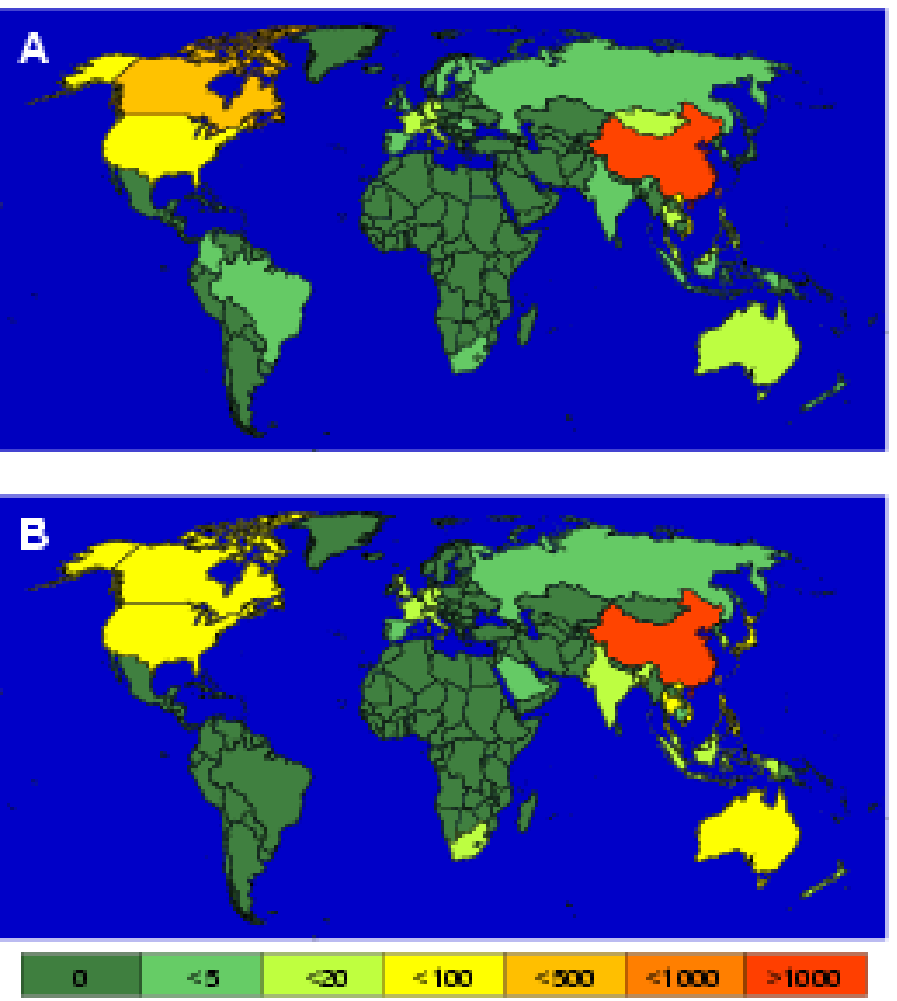}\end{center}

\caption{\label{cap:fig2}\textsf{\textbf{(a)}} \textsf{Geographical representation
of the global spreading of probable SARS cases on May 30th, 2003 as
reported by the WHO and CDC. The first cases of SARS emerged in mid
November 2002 in Guangdong Province, China\cite{MMWRww}. The disease
was then carried to Hong Kong on the February 21st, 2003 and began
spreading around the world along international air travel routes,
as tourists and the medical doctors who treated the early cases traveled
internationally. As the disease moved out of southern China, the first
hot zones of SARS were Hong Kong, Singapore, Hanoi (Viet Nam) and
Toronto (Canada), but soon cases in Taiwan, Thailand, the United States,
Europe and elsewhere were reported.} \textsf{\textbf{(a)}} \textsf{Geographical
representation of the results of our simulations 90 days after an
initial infection in Hong Kong, The simulation corresponds to the
real SARS infection at the end of May, 2003. Since our Since the simulations
cannot describe the infection in China, where the disease started
in November 2002, we used the WHO data for China.}}
\end{figure}

Fig.~\ref{cap:fig2} depicts a geographical representation of the
results of our simulations. Initially, an infected individual was
placed in Hong Kong. For this initial condition we simulated $1000$
realizations of the stochastic model and computed the mean value $\left\langle I(t)\right\rangle $
of the number of infecteds at each node $i=1,...,M$ of the network.
Since the size of catchment areas varies on many scales, the fluctuation
range is best quantied by the means of the relative variance of $z=\log\, I$,
i.e. $\eta=\sqrt{\left\langle z^{2}\right\rangle -\left\langle z\right\rangle ^{2}}/<z>.$
In our simulations we computed this measure for every $i$ of the
network. Fig~\ref{cap:fig2} shows the prediction of our model for
the spread of SARS at $t=90$ days after the initial outbreak in Hong
Kong (February 19, 2003), corresponding to the May 20, 2003. The results
of our simulations are in remarkable agreement with the worldwide
spreading of SARS as reported by the WHO (compare Fig.~\ref{cap:fig2}):
There is an almost one-to-one correspondence between infected countries
as predicted by the simulations and the WHO data. 

\begin{table}
\begin{center}\begin{tabular}{|c|c|c|c|c|c|c|}
\hline 
Country&
WHO&
WHO &
\multicolumn{4}{c|}{Simulation}\tabularnewline
&
05/20/2003&
05/30/2003&
Average&
$\eta$&
Min&
Max\tabularnewline
\hline 
Hong Kong&
1718&
1739&
1951&
0.35&
1373.9&
2770.4\tabularnewline
\hline
Taiwan&
383&
676&
318.2&
0.55&
184.0&
550.3\tabularnewline
\hline 
Singapore&
206&
206&
136.6&
0.68&
69.4&
268.7\tabularnewline
\hline 
Japan&
-&
-&
60.4&
0.84&
26.6&
137.0\tabularnewline
\hline 
Canada&
140&
188&
41.8&
0.94&
16.4&
106.6\tabularnewline
\hline 
USA&
67&
66&
65.9&
0.84&
28.4&
152.7\tabularnewline
\hline 
Vietnam&
63&
63&
49.2&
0.86&
20.7&
116.3\tabularnewline
\hline 
Philippines&
12&
12&
30.0&
0.97&
6.2&
50.7\tabularnewline
\hline 
Germany&
9&
10&
14.4&
1.1&
4.8&
43.1\tabularnewline
\hline 
Netherlands&
-&
-&
5.9&
1.09&
2.0&
17.6\tabularnewline
\hline 
Bangladesh&
-&
-&
10&
1.15&
3.2&
31.6\tabularnewline
\hline 
Mongolia&
9&
9&
-&
-&
-&
-\tabularnewline
\hline 
Italy&
9&
9&
5.3&
1.02&
1.9&
14.6\tabularnewline
\hline 
Thailand&
8&
8&
35.4&
0.89&
14.5&
86.8\tabularnewline
\hline 
France&
7&
7&
7.6&
1.09&
2.6&
22.6\tabularnewline
\hline 
Australia&
6&
6&
27.0&
1.05&
10.1&
72.5\tabularnewline
\hline 
Malaysia&
7&
5&
17.7&
1.05&
6.2&
50.7\tabularnewline
\hline 
United Kingdom&
4&
4&
16.7&
1.04&
5.9&
47.0\tabularnewline
\hline
\end{tabular}\end{center}

\caption{\label{cap:tab1}\textsf{A comparison of the SARS case reports provided
by the WHO and the results of our simulation for all countries with
a reported case number $\ge4$. The expected number of infecteds predicted
by our model is estimated by the average over $1000$ realizations
of the stochastic model. The epidemic was simulation for $t=90$ days
after the initial outbreak of SARS in Hong Kong (February 19, 2003),
yielding a simulation end of May 20, 2003. The range defined by column
$6$ and $7$ was computed by means of the fluctuation measure $\eta$
(see text) which is approximately one for all countries.}}
\end{table}

Also the orders of magnitude of the numbers of infected individuals
in a country agree (Table~\ref{cap:tab1}). While for most countries
the reported cases by the WHO lie within the fluctuation range, two
deviation between the reported cases and the predictions of the simulation
are apparent: Our simulations predict a relatively high number of
SARS cases in Japan (between 26.6 and 137.0). However, the Japanese
Government reported no confirmed case (only 5 suspected cases) of
SARS in Japan, as of May 30, 2003. How a single realization may deviate
from the expectation can be seen from the difference between the simulation
and the reported cases in the USA and Canada. The simulations show
that on average the USA should have a higher number of SARS cases
than Canada, although the opposite was reported by the WHO. The impact
of the inherent stochasticity of the infection and traveling dynamics
is discussed in the next section.

\section{The impact of fluctuations}

\begin{figure}
\includegraphics[%
  width=1.0\columnwidth]{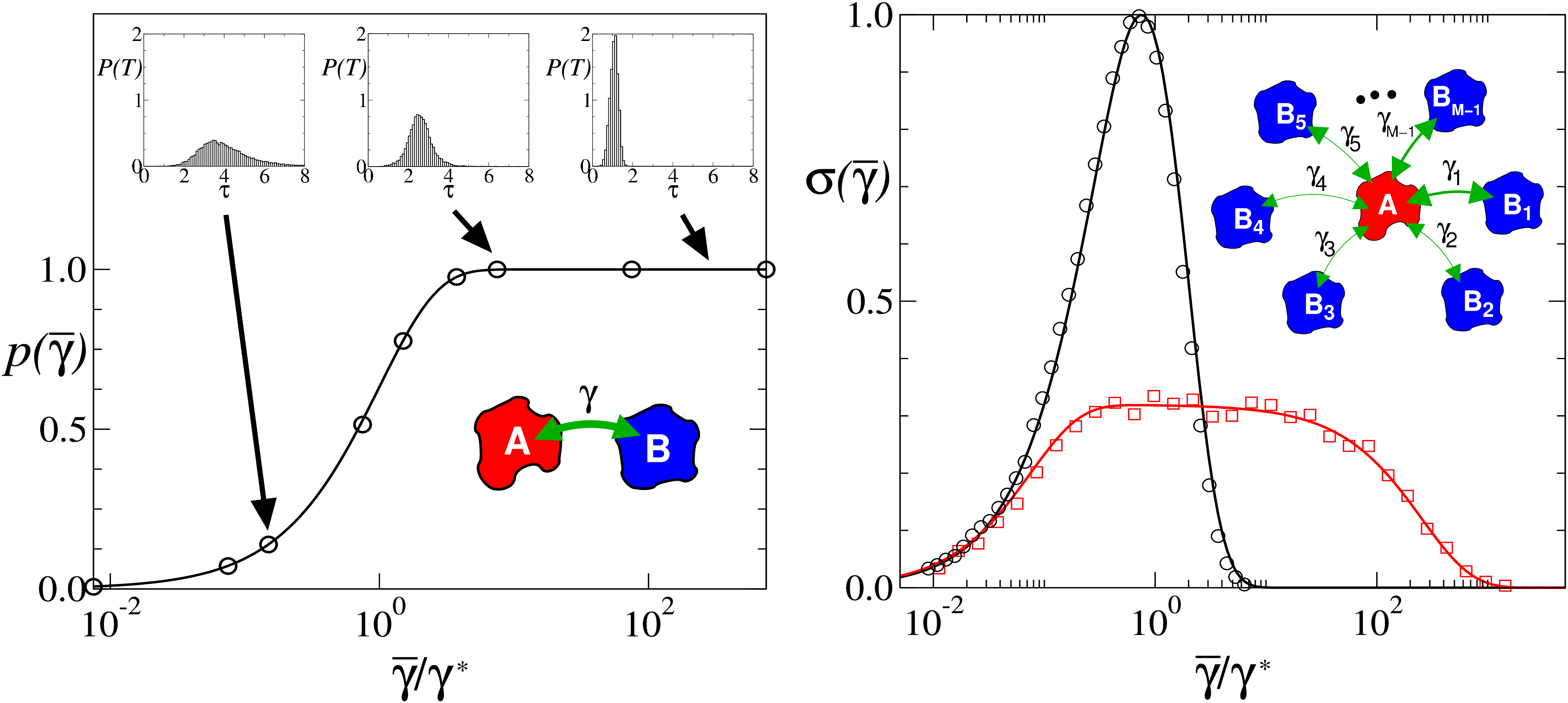}

\caption{\label{cap:fig3}\textsf{Two confined populations with exchange of
individuals. In each population the dynamics is governed by the SIR-reaction
scheme~(\ref{eq:pups}). Individuals travel from on population to
the other at a rate $\gamma$. Parameters are $N_{A}=N_{B}=10,000$,
$R_{0}=4$ and an initial number of infecteds $I_{0}=20$ in population
$A$} \textsf{\textbf{Left:}} \textsf{The probability $p(\gamma)$
of an outbreak occurring in population $B$ as a function of transition
rate $\gamma$. The insets depict histograms of the time lag $T$
between the outbreaks in $A$ and $B$ for those realization for which
an outbreak occurs in $B$. The circles are results of the simulations
of $100,000$ realizations, the solid curve is the analytic result
of Eq.~(\ref{eq:threshold})} \textsf{\textbf{Right:}} \textsf{A
star-shaped network with a central population $A$ connected to $M-1$
populations $B_{1},...,B_{M-1}$with rates $\gamma_{1},...,\gamma_{M-1}$.
The cumulated variance (Eq.~\ref{eq:typicalvariance}) for a star
network with 32 populations is depicted as a function of the average
transmission rate $\bar{\gamma}$. Two cases are exemplified: equal
rates (circles) and distributed rates according to Eq.~\ref{eq:cooledist}
with $\gamma_{{\rm max}}/\gamma_{{\rm min}}\approx1000$ (squares).
The solid lines show the analytical results given by Eq.~\ref{eq:sigmadef}
and Eq.~\ref{eq:threshold}. Parameters are $N_{A}=N_{B}=10,000$,
$R_{0}=4$ and an initial number of infecteds $I_{0}=20$ in population
$A$. The numerical values are obtained by calculating the variance
of the fluctuations of 100 different realizations of the epidemic
outbreak for each $\bar{\gamma}$.}}
\end{figure}

Bearing in mind the low number of infections and the small value of
$\rho_{0}$ for SARS, the high degree of predictability, i.e. the
low impact of fluctuations on the network level, is rather surprising,
especially because our simulations take into account the full spectrum
of fluctuations of disease transmission, recovery and dispersal and
that the system evolves on a highly complex network. Naively, one
expects that dispersal fluctuations between two given populations
are amplified as the epidemic spreads globally and that no prediction
can be made. In order to clarify this important point, consider the
system of two confined populations $A$ and $B$ which exchange individuals
as depicted in Fig.~\ref{cap:fig3}. For simplicity we assume that
both populations have the same size (i.e. $N_{A}=N_{B}=N$) and individuals
traverse at a rate $\gamma$. Now assume that initially a small number
of infected $I_{0}$ is introduced to population $A$ without any
infecteds contained in $B$. For a sufficiently high number of infecteds
in $A$ an epidemic occurs. For $\gamma>0$ infecteds are introduced
to $B$ and a subsequent outbreak may occur in $B$ after a time lag
$T$. Fig.~\ref{cap:fig3} depicts the results of simulations for
two populations with $N=10,000$ and $\rho_{0}=4$. Various realizations
of the time course $I_{A}(t)$ and $I_{B}(t)$ of the epidemic in
both populations we computed. The initial number of infecteds in population
was $I_{A}(t=0)=I_{0}=20$. The left panel depicts the probability
$p(\gamma)$ of an outbreak occuring in population $B$ as a function
if the transition rate $\gamma$. For large enough rates the probability
is nearly unity, since a sufficient number of infecteds is introduced
to $B$. For very low rates $\gamma$ no infecteds are introduced
to $B$ during the time span of the epidemic in $A$ and thus $p(\gamma)\rightarrow0$
as $\gamma\rightarrow0$. For intermediate values of $\gamma$ the
probability $p(\gamma)$ is neither one nor unity and the time course
in population $B$ cannot be predicted with certainty. The function
$p(\gamma)$ is given by\begin{equation}
p(\gamma)=1-\exp(-\gamma/\gamma^{\star}),\label{eq:threshold}\end{equation}
where the critical rate $\gamma^{\star}$ is a function of the parameters
$\rho_{0}$ and $N$. The insets depict histograms of the time lag
$T$ for those realization for which an outbreak occured in $B$.
Each histogram corresponds to a different transition rate $\gamma$.
The smaller $\gamma$ the higher the variability in $T$. Note that
even in a range in which $p(\gamma)\approx1$, the time lag $T$ is
still a stochastic quantity with a high degree of variance (see also
the supplement material).

Consequently, the introduction of stochastic exchange of infected
individuals leads to a lack of predictability in the time of onset
of the initially uninfected population. In the light of the analysis
of two populations, the predictability in the case of SARS on the
aviation network seems even more puzzling.

The situation changes drastically in networks which exhibits a high
degree of variability in the rate matrix $\gamma_{ij}$. Clearly,
this is the case for the aviation network. Consider the simple network
depicted in Fig.~\ref{cap:fig3}. Each population contains $N$ individuals.
A central population $A$ is coupled to a set of $M-1$ surrounding
populations $B_{1},...B_{M-1}$. Assume that initially a number of
infecteds $I_{0}$ is introduced to the central population $A$ such
that an outbreak occurs. The entire set of rates $\left\{ \gamma_{j}\right\} {}_{j=1,...,M-1}$
determines the behaviour in the surrounding populations. If all rates
$\gamma_{j}$ are identical and very small we expect no infection
to occur in the $B_{j}$, for large enough $\gamma_{j}$ an outbreak
will occur in every $B_{j}$. In the aviation network, however, transition
rates are distributed on many scales and the response of the network
to a central outbreak depends on the statistical properties of this
distribution denoted by$q(\gamma)$. In order to quantify the reaction
of the network we introduce for each surrounding population a binary
number $\xi_{j}$ with $j=1,...,M-1$ which is unity if an outbreak
occurs in $B_{j}$ and zero if it doesn't. According to Eq.~(\ref{eq:threshold})
for a given rate $\gamma$ this quantity is a random number with a
conditional probability density $p(\xi_{i}|\gamma)=(1-p(\gamma))\,\delta(\xi_{i})+p(\gamma)\,\delta(\xi_{i}-1)$.
The variability of the network is thence quantified be the cumulative
variance per population and we define\begin{equation}
\sigma=\frac{4}{M-1}\sum_{i}\text{var}(\xi_{i})=\int\diff{\gamma}\, p(\gamma)\,(1-p(\gamma))\, q(\gamma)\label{eq:sigmadef}\end{equation}
as a measure for the uncertainty of the network response. If for example
$q(\gamma)=\delta(\gamma-\bar{\gamma})$, i.e. all transition rate
are identical and equal to $\bar{\gamma}$, then $\sigma(\bar{\gamma})=4\, p(\bar{\gamma})(1-p(\bar{\gamma})$,
which is unity for $p(\bar{\gamma})=1/2$. Comparing with Eq.~(\ref{eq:threshold})
we see that when $\bar{\gamma=\gamma^{\star}\log2}$ the system with
identical transition rates $\gamma_{i}=\bar{\gamma}$ exhibits the
highest degree of unpredictability when the rates are of the order
of the critical rate defined by~(\ref{eq:threshold}). The function
$\sigma(\bar{\gamma})$ is shown in Fig.~\ref{cap:fig3}. 

Now assume that the rate $\gamma_{j}$ are drawn from a distribution\begin{equation}
q(\gamma)=\frac{1}{\log(\gamma_{\text{max}}/\gamma_{\min})}\frac{1}{\gamma}\qquad\gamma_{\text{max}}\leq\gamma\leq\gamma_{\min}.\label{eq:cooledist}\end{equation}
which implies a high degree of variance within the interval $\left[\gamma_{\text{min}},\gamma_{\text{max}}\right]$
(i.e. $\gamma_{j}$ is distributed uniformly on a logarithmic scale).
This high variability in rates drastically changes the predictability
of the system. Inserting into Eq.~(\ref{eq:sigmadef}) yields $\sigma(\bar{\gamma})$
for strongly distributed rates. In Fig.~\ref{cap:fig3} this function
is compared to a system of identical transition rates. On one hand,
for intermediate values of $\gamma\approx\gamma^{\star}$ the predictability
is much higher than in the system of identical rates. This is a rather
counterintuitive result. Despite the additional randomness in transition
rates, the degree of determinism is increased.

\section{Control Strategies}

\begin{figure}
\includegraphics[%
  width=1.0\columnwidth]{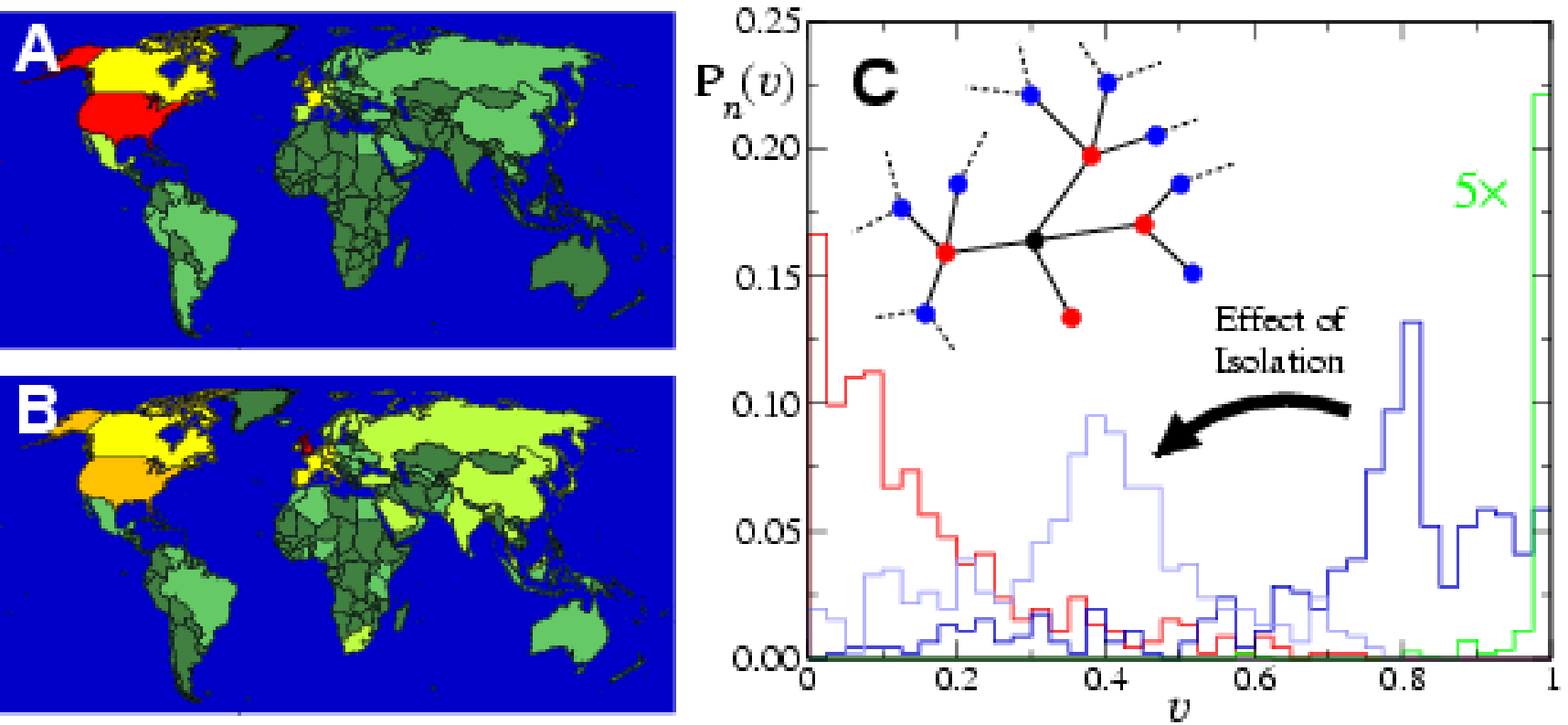}

\caption{\label{cap:fig4}\textsf{\textbf{Left:}} \textsf{Geographical representation
of the results two simulations of hypothetical SARS outbreaks 90 days
after an initial infection in (a) New York and (b) London for the
same parameters and color code as in Fig.~\ref{cap:fig2}.} \textsf{\textbf{Right:}}
\textsf{Impact and control of epidemics. The probability $p_{n}(v)$
of having to vaccinate a fraction $v$ of the population in order
to prevent the epidemic from spreading, if an initial infected individual
is permitted to travel $n=1$ (red), $2$ (blue), and $3$ (green)
times. The probability $p_{n}(v)$ is estimated by placing the infected
individual on a node $i$ (black dot) of the network. The fraction
$v_{i}$ associated with node $i$ is given by the number of susceptibles
in a subnetwork which can be reached by the infected individual after
$n=1,\,2$ and $3$ steps. Histogramming $v_{i}$ for all nodes $i$
yields an estimate for $p_{n}(v)$. The light-blue curve depicts the
strong impact of isolating only $2\%$ of the largest cities after
an initial outbreak ($n=2$) and is to be compared to the blue curve.}}
\end{figure}

Fig.~\ref{cap:fig4} exemplify how our model can be employed to predict
endangered regions if the origin of a future epidemic is located quickly.
The figure depict simulations of the global spread of SARS at $t=90$
days after hypothetical outbreaks in New York and London, respectively.
Despite the worldwide spread of the epidemic in each case, the degree
of infection of each country differs considerably, which has important
consequences for control strategies.

Vaccination of a fraction of the population reduces the fraction of
susceptibles and thus yields a smaller effective reproduction number
$\rho$. If a sufficiently large fraction is vaccinated, $\rho$ drops
below 1 and the epidemic becomes extinct. The global aviation network
can be employed to estimate the fraction of the global population
that needs to be vaccinated in order to prevent the epidemic from
spreading. Fig.~\ref{cap:fig4} demonstrates that a quick response
to an initial outbreak is necessary if global vaccination is to be
avoided. The Figure depicts the probability $p_{n}(v)$ of having
to vaccinate a fraction $v$ of the population if an infected individual
is randomly placed in one of the cities and permitted to travel $n=1,\,2$
or $3$ times. For the majority of originating cities the initial
spread is regionally confined and thus a quick response to an outbreak
requires only a vaccination of a small fraction of the population.
However, if the infected individual travels twice, the expected fraction
$\left\langle v\right\rangle $ of the population which needs to be
vaccinated is considerable ($74.58\%$). For $n=3$ global vaccination
is necessary. 

As a reaction to a new epidemic outbreak, it might be advantageous
to impose travel restrictions to inhibit the spread. Here we compare
two strategies: (i) the shutdown of individual connections and (ii)
the isolations of cities. Our simulations show that an isolation of
only $2\%$ of the largest cities already drastically reduces $\left\langle v\right\rangle $
(with $n=2$) from $74.58\%$ to $37.50\%$ (compare the blue and
light-blue curves in Fig.~\ref{cap:fig4}). In contrast, a shutdown
of the strongest connections in the network is not nearly as effective.
In order to obtain a similar reduction of $\left\langle v\right\rangle $
the top $27.5\%$ of connections would need to be taken off the network.
Thus, our analysis shows that a remarkable success is guaranteed if
the largest cities are isolated as a response to an outbreak.

In a globalized world with millions of passengers traveling around
the world week by week infectious diseases may spread rapidly around
the world. We believe that a detailed analysis of the aviation network
represents a cornerstone for the development of efficient quarantine
strategies to prevent diseases from spreading. As our model is based
on a microscopic description of traveling individuals our approach
may be considered a reference point for the development and simulation
of control strategies for future epidemics.

We thank E.~Bodenschatz for critical reading of the manuscript and
stimulating discussions. This research was supported in part by the
National Science Foundation under Grant No. PHY99-07949.

\bibliographystyle{apsrev}
\bibliography{./simon,./paper_local}

\end{document}